\documentclass[11pt, A4]{article}

\usepackage{graphicx}
\usepackage{epstopdf}
\usepackage{amsmath}
\usepackage{amssymb}
\usepackage{amsfonts}
\usepackage{amsthm}
\usepackage[usenames]{color}
\usepackage{array}

\definecolor{GC}{rgb}{0,0.0,0.65}
\usepackage[
      colorlinks=true,
      linkcolor=blue,
      urlcolor=blue,
      filecolor=blue,
      citecolor=red,
      pdfstartview=FitV,
      pdftitle={},
      pdfauthor={},
      pdfsubject={},
      pdfkeywords={},
      pdfpagemode=None,
      bookmarksopen=true
]{hyperref}

\usepackage{epsfig}

\usepackage{hyperref}

\def\beq{\begin{eqnarray}}
\def\eeq{\end{eqnarray}}

\def\be{\begin{equation}}
\def\ee{\end{equation}}
\def\bea{\begin{eqnarray}}
\def\eea{\end{eqnarray}}

\newcommand{\rom}[1]{\mathrm{#1}}

\def\nn{\nonumber}

\begin{document}
\pagestyle{myheadings}
\markboth{\textsc{\small }}{%
  \textsc{\small Inverse Scattering Construction of a Dipole Black Ring}} \addtolength{\headsep}{4pt}

%% \begin{titlepage}
\begin{flushright}
\texttt{AEI-2011-056}\\
\texttt{ULB-TH/11-19}
\end{flushright}
\vspace{1cm}

\begin{centering}

  \vspace{0cm}

  \textbf{\Large{Inverse Scattering Construction of a Dipole Black Ring}}

  \vspace{0.8cm}

  {\large Jorge V. Rocha$^\natural$, Maria J. Rodriguez$^\sharp$, and Amitabh Virmani$^\Diamond$ }

  \vspace{0.5cm}

\begin{minipage}{.9\textwidth}\small \it \begin{center}
$\natural$
Centro Multidisciplinar de Astrof\'isica - CENTRA,\\
Dept. de F\'isica, Instituto Superior T\'ecnico, Technical University of Lisbon,\\
Av. Rovisco Pais 1, 1049-001 Lisboa, Portugal \\
 {\tt jorge.v.rocha@ist.utl.pt}\\
$ \, $ \\

$\sharp$
Max-Planck-Institut f\"ur Gravitationsphysik,\\
Albert-Einstein-Institut, 14476 Golm, Germany \\
{\tt maria.rodriguez@aei.mpg.de}
\\ $ \, $ \\

$\Diamond$
Physique Th\'eorique et Math\'ematique,  \\ Universit\'e Libre de
Bruxelles and International Solvay Institutes\\ Campus
Plaine C.P. 231, B-1050 Bruxelles,  Belgium\\
{\tt avirmani@ulb.ac.be}
    \end{center}
\end{minipage}

\end{centering}

\vspace{1cm}

%\begin{center}
%  \begin{minipage}{.9\textwidth}
%    \begin{center}

\begin{abstract}

Using the inverse scattering method in six dimensions we construct the dipole black ring of five dimensional Einstein-Maxwell-dilaton theory with dilaton coupling $a = 2 \sqrt{2/3}$. The 5d theory can be thought of as the NS sector of low energy string theory in Einstein frame.  It can also be obtained by dimensionally reducing six-dimensional vacuum gravity on a circle. Our new approach uses GL(4,~R) integrability structure of the theory inherited from six-dimensional vacuum gravity. Our approach is also general enough to potentially generate dipole black objects carrying multiple rotations as well as more exotic multi-horizon configurations.

\end{abstract}
\vfill

%\noindent \mbox{}
%\raisebox{-3\baselineskip}{%
%  \parbox{\textwidth}{ \mbox{}\hrulefill\\[-4pt]}}

\thispagestyle{empty} \newpage

\tableofcontents

\setcounter{equation}{0}

%%%%%%%%%%%%%%%%%%%%%%%%%%%%%%%%%%%%%%%
\section{Introduction}
Almost a decade after the first black ring solution was written down by Emparan and Reall~\cite{Emparan:2001wn}, black rings are still a continual source of excitement in higher dimensional general relativity and in microscopic description of black holes in string theory; see~\cite{Emparan:2006mm,Emparan:2008eg,Rodriguez:2010zw} for reviews. A technical problem in the black ring literature, whose solution is still elusive, is the construction of the most general asymptotically flat black ring in a simple five-dimensional supergravity theory --- a five-parameter black ring solution carrying mass, two angular momenta, electric charge and magnetic dipole charge. Not only this solution is not known, but also the requisite set of techniques to carry out a systematic construction are not fully understood. The most important stumbling block in these considerations is the dipole charge. The original dipole solution~\cite{Emparan:2004wy} was constructed using educated guesswork. There also exists in the literature an algorithmic construction of a black ring solution carrying dipole charge~\cite{Yazadjiev:2006ew}. However, it cannot be employed to generate multiple rotations. In this paper we remedy this situation for Einstein-Maxwell-dilaton theory with dilaton coupling $a = 2 \sqrt{2/3}$. This theory is obtained by dimensionally reducing six-dimensional vacuum gravity on a circle. It can also be thought of as the NS sector of low energy string theory in Einstein frame. We exploit GL(4, R) integrability structure of the theory inherited from six-dimensional vacuum gravity and thus employ the inverse scattering method in six dimensions. A key point in this procedure is the identification of the seed that can be used to generate the dipole charge.

The new construction presented here below can in principle be used to construct the doubly spinning dipole black ring and/or the most general black ring in this theory. This situation is in contrast with the previous approaches, e.g.,~\cite{Yazadjiev:2006ew}, which rely on a specific SL(2, R) $\times$ SL(2, R) structure, and cannot be extended in a natural way to construct, say, the doubly spinning dipole black ring. Our approach does not rely on SL(2, R) $\times$ SL(2, R) structure, but rather uses the full GL(4, R) symmetry of the theory. For a discussion on using integrability techniques to construct solutions of supergravity theories we refer the reader to~\cite{Figueras:2009mc, Alekseev:2010mx}.
In the next section we present our construction; in the following section we conclude with some observations and comments.

%%%%%%%%%%%%%%%%%%%%%%%%%%%%%%%%%%%%%%%
\section{Inverse scattering construction of a dipole black ring}
This section is the core of the paper. We first present our setup and then the construction of the dipole ring solution.

%%%%%%%%%%%%%%%%%%%%%%%%%%%%%%%%%%%%%%%
\subsection{The set-up}
The action of the five-dimensional theory we are interested in is
\be
S = \frac{1}{16 \pi G} \int d^5 x \sqrt{-g} \left(R - \frac{1}{2} \partial_\mu \phi \partial^{\mu}\phi - \frac{1}{4} e^{-a \phi}F_{\mu \nu} F^{\mu \nu} \right),
\label{action}
\ee
with
\be
a = \frac{2 \sqrt{2}}{\sqrt{3}}.
\ee
This action can also be obtained by the circle reduction of six-dimensional vacuum gravity using the ansatz
\be
ds^2_{6} = e^{\frac{\phi}{\sqrt{6}}} ds^2_5 + e^{-\frac{\sqrt{3}\phi}{\sqrt{2}}} (dw + A)^2.
\label{metric6d}
\ee
The sixth dimension is parametrized by $w$ and $F = dA$. The five-dimensional theory naturally supports a magnetic one-brane, an electric zero-brane (and smeared versions thereof), and dipole black rings.  In the literature solutions of the `electric' version of the theory under consideration are studied in greater detail~\cite{Harmark:1999xt}. This is obtained by defining
\be
H = e^{-a\phi} \star_5 F, \qquad H = dB,
\ee
then
\be
S = \frac{1}{16 \pi G} \int d^5 x \sqrt{-g} \left(R - \frac{1}{2} \partial_\mu \phi \partial^{\mu}\phi - \frac{1}{12} e^{a \phi}H_{\mu \nu \rho} H^{\mu \nu \rho} \right).
\label{actionNS}
\ee
The action \eqref{actionNS} can be interpreted as the NS sector of low energy string theory in Einstein frame. The fundamental string is a solution of this action. The dilaton $\sigma$ of string theory is $\sigma = -\sqrt{\frac{3}{8}}\phi$ and the string metric is $g^{\rom{(s)}}_{\mu \nu} = e^{-\sqrt{\frac{2}{3}} \phi} g_{\mu \nu}$. Via string dualities solutions of \eqref{actionNS} can be related to other single brane configurations.

In the following we restrict our attention to the `magnetic' version \eqref{action} and its uplift to six-dimensional vacuum gravity via \eqref{metric6d}.

%%%%%%%%%%%%%%%%%%%%%%%%%%%%%%%%%%%%%%%
\subsection{Seed and soliton transformations}
The inverse scattering method has been widely used to construct analytically many of currently available asymptotically flat black hole solutions in four and five dimensions. The construction consists in a series of soliton transformations on a seed solution. Through these `dressing' transformations, or Belinsky-Zakharov (BZ) procedure~\cite{BZ}, a more general solution can be generated exhibiting new free parameters.  We use this method below. For our analysis we also make use of the results of higher-dimensional Weyl and rod diagram representations from~\cite{Emparan:2001wk, Harmark:2004rm, Kleihaus:2010pr}.

\begin{figure}[t!]
\centering{
\includegraphics{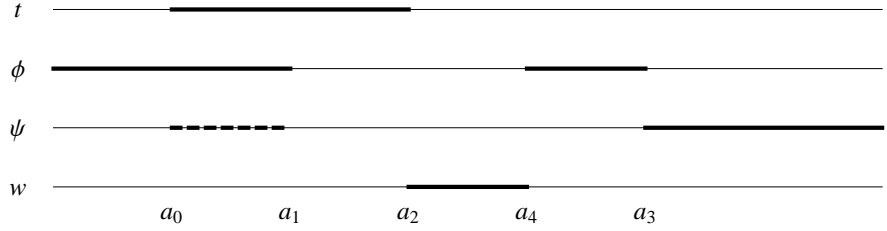}}
%\bigskip
\caption{The figure depicts sources for the seed metric $G_0$.  The rods
  are located in the $(\rho, z)$ space at the $z$-axis with $\rho=0$. The solid rods have
  positive density and the dashed rod has negative density. They add up to an infinite
  rod with uniform density such that $\det G_0 = -\rho^2$.}
\label{roddiagram}
\end{figure}

The rod configuration for the seed solution of dipole black ring when lifted to six-dimensions is shown in figure~\ref{roddiagram}. The thick solid lines correspond to rod sources of uniform density +1/2, and the dashed line corresponds to a rod source of uniform density $-$1/2. The rod in the $t$ direction corresponds to the horizon. In figure~\ref{roddiagram} when $a_0 = a_1$ and $a_4 = a_2$ the negative density rod and the rod in the $w$ direction are eliminated and the solution describes a neutral static black ring multiplied with the flat direction $w$. This is an unbalanced configuration. The negative density rod is included in the seed following~\cite{Elvang:2007rd} to facilitate adding the $S^1$ angular momentum to the ring. The positive density rod $[a_2,a_4]$ along the $w$ direction is included to facilitate adding dipole charge to the ring.

The seed metric corresponding to the rod configuration of figure~\ref{roddiagram} is given by
\be
ds^2_6 = (G_0)_{ab}\; dx^a dx^b+ e^{2\nu_0}(d\rho^2+dz^2),
\label{confmetric}
\ee
where
\be
G_0 = \verb+diag+\left\{ -\frac{\mu_0}{\mu_2}, \frac{\rho^2 \mu_4}{\mu_1 \mu_3}, \frac{\mu_1 \mu_3}{\mu_0}, \frac{\mu_2}{\mu_4} \right\}, \qquad \det G_0 = - \rho^2,
\label{seed}
\ee
and the conformal factor of the seed is
\be
e^{2\nu_0} = k^2 \frac{\mu_1 \mu_3}{\mu_0} \frac{(\mu_0 \mu_1+\rho^2)(\mu_0 \mu_2+\rho^2)(\mu_0 \mu_3+\rho^2)(\mu_1 \mu_4+\rho^2)(\mu_2 \mu_4+\rho^2)(\mu_3 \mu_4+\rho^2)}{(\mu_1 \mu_3+\rho^2)^2 \prod_{i=0}^{4}(\mu_i^2+\rho^2)}.
\label{e2nu}
\ee
The integration constant $k$ will be fixed in the next subsection. Our ordering of coordinates is $ x^a=(t, \phi, \psi, w)$, with $t$ corresponding to the timelike coordinate, $\phi$ describing the azimuthal angle on the $S^2$ and $\psi$ being the angle along the $S^1$ component of the ring. The $\mu_i$'s are pole trajectories of the dressing matrices, commonly referred to as solitons, $\mu_i = \sqrt{\rho^2 + (z-a_i)^2} - (z-a_i)$, and $a_i$ are the rod endpoints. In writing these expressions we have followed established conventions from the literature.  We refer the reader to e.g. \cite{Emparan:2008eg} for notational details.

We assume the ordering
\be
a_0 \le a_1 \le a_2 \le a_4 \le a_3. \label{ordering}
\ee
The labeling (and hence the ordering) is a little unusual, but is motivated to simplify the presentation of the solution after the inverse scattering transformation. We will see that the endpoints $a_0$ and $a_4$ will not appear in the rod diagram of the final solution.

The solution \eqref{seed} and \eqref{e2nu} with ordering \eqref{ordering} is singular and not in itself of direct physical interest. First, with a 1-soliton transformation we add an anti-soliton which mixes $t$ and $\psi$ directions in such a way that the negative density rod $[a_0,a_1]$ moves in the $t$ direction and cancels the segment $[a_0,a_1]$ of the positive density rod. Taking $a_4 =a_2$ at this stage gives the singly spinning black ring of~\cite{Emparan:2001wn}; this step is the six dimensional analogue of the one used in \cite{Elvang:2007rd}. Next, keeping $a_2 \le a_4$, we apply a further 1-soliton transformation that gives the dipole black ring. This latter 1-soliton transformation mixes the $w$ and $\phi$ directions in such a way that the positive density rod $[a_2,a_4]$ moves in the $\phi$ direction and \emph{merges} with the positive density rod $[a_4, a_3]$. The first transformation leaves behind naked singularities at $z=a_0$, but choosing the corresponding BZ vector appropriately completely eliminates this singularity. Requiring the final rod $[a_2, a_3]$ to have uniform rod direction fixes the second BZ vector.

In more detail the steps for generating the dipole ring solution by a 2-soliton transformation are as follows:
\begin{enumerate}
\item Perform two 1-soliton transformations on the seed solution \eqref{seed} followed by a rescaling. More precisely
$(i)$ remove an anti-soliton at $z=a_0$ with trivial BZ vector $(1,0,0,0)$,
$(ii)$ remove a soliton at $z=a_4$ with trivial BZ vector $(0,0,0,1)$,
$(iii)$ rescale the metric by $\mu_0/\mu_4$.
The resulting metric is
\be
G_0' = \verb+diag+\left\{ -\frac{\bar{\mu}_4}{\mu_2}, -\frac{\mu_0 \bar{\mu}_3}{\mu_1}, \frac{\mu_1 \bar{\mu}_4}{\bar{\mu}_3}, \frac{\mu_0}{\bar{\mu}_2} \right\}.
\label{seed1}
\ee
This metric will be the seed for our next soliton transformation.
\item The generating matrix corresponding to \eqref{seed1} is:
\be
\psi_0'(\lambda,\rho, z) = \verb+diag+\left\{ -\frac{\bar{\mu}_4 - \lambda}{\mu_2 -\lambda}, -\frac{(\mu_0 -\lambda) (\bar{\mu}_3-\lambda)}{\mu_1-\lambda}, \frac{(\mu_1 -\lambda)( \bar{\mu}_4 - \lambda)}{\bar{\mu}_3 -\lambda}, \frac{\mu_0 - \lambda}{\bar{\mu}_2 - \lambda} \right\}.
\label{Gseed1}
\ee
\item Perform now a 2-soliton transformation with $G_0'$ as seed and undo the rescaling. More precisely
$(i)$ add an anti-soliton at $z=a_0$ with BZ vector $(1,0,c_1,0)$,
$(ii)$ add a soliton at $z=a_4$ with BZ vector $(0,c_2,0,1)$, and 
$(iii)$ rescale by $\mu_4/\mu_0$.
Denote the final metric by $G$. The final rescaling ensures that $\det G = - \rho^2$.
\item Construct $e^{2\nu}$. The result $(e^{2\nu}, G)$ is the six-dimensional solution we want. Appropriately tuning $c_1$ and $c_2$ and Kaluza-Klein reducing this solution along the $w$ direction we obtain the smooth five-dimensional dipole black ring solution of the theory under consideration.
\end{enumerate}
In the next subsection we first present the resulting solution in the Weyl coordinates and analyze the rod structure and regularity of the solution. We then transform it into the more convenient $(x,y)$ coordinates. The resulting metric is shown to be identical to the one presented in~\cite{Emparan:2004wy}.

%%%%%%%%%%%%%%%%%%%%%%%%%%%%%%%%%%%%%%%
\subsection{Dipole black ring solution}
The $G_{tt}$, $G_{t\psi}$ and $G_{\psi \psi}$ components of the metric after performing the above described inverse scattering transformations take the form:
\bea
G_{tt} &=& -\frac{\mu_0 \Big[  \mu_3 (\mu_0 \mu_1 + \rho^2)^2 (\mu_0 \mu_2 + \rho^2)^2 - c_1^2 \mu_1 \mu_2 (\mu_0-\mu_3)^2 \rho^4 \Big]}{\mu_2 \Big[\mu_3 (\mu_0 \mu_1 + \rho^2)^2 (\mu_0 \mu_2 + \rho^2)^2 + c_1^2 \mu_0^2 \mu_1 \mu_2 (\mu_0-\mu_3)^2 \rho^2 \Big]},\label{metric6dWeyl1}\\
G_{t\psi} &=& - \frac{c_1 \mu_1 \mu_3 (\mu_0 - \mu_3)  (\mu_0^2 + \rho^2)(\mu_0 \mu_1 + \rho^2)(\mu_0 \mu_2 + \rho^2)}{\Big[\mu_3 (\mu_0 \mu_1 + \rho^2)^2(\mu_0 \mu_2 + \rho^2)^2 + c_1^2 \mu_0^2 \mu_1 \mu_2 (\mu_0 -\mu_3)^2 \rho^2\Big]}, \label{metric6dWeyl2}\\
G_{\psi \psi} &=& \frac{\mu_1 \mu_3 \Big[ \mu_3  (\mu_0 \mu_1 + \rho^2)^2(\mu_0 \mu_2 + \rho^2)^2 - c_1^2 \mu_0^4 \mu_1 \mu_2 (\mu_0 - \mu_3)^2 \Big]}{\mu_0 \Big[\mu_3 (\mu_0 \mu_1 + \rho^2)^2(\mu_0 \mu_2 + \rho^2)^2 + c_1^2 \mu_0^2 \mu_1 \mu_2 (\mu_0 -\mu_3)^2 \rho^2\Big]}\label{metric6dWeyl3}.
\eea
Similarly, the $G_{\phi \phi}$, $G_{w\phi}$ and $G_{ww}$ components of the metric take the form:
\bea
G_{\phi \phi} &=& \frac{ \rho^2\Big[ \mu_1 \mu_4^2  (\mu_2 \mu_4 + \rho^2)^2 (\mu_3 \mu_4 + \rho^2)^2 + c_2^2 \mu_2 \mu_3 (\mu_1-\mu_4)^2 \rho^6 \Big]}{\mu_1 \mu_3 \mu_4 \Big[\mu_1 (\mu_2 \mu_4 + \rho^2)^2 (\mu_3 \mu_4 + \rho^2)^2 - c_2^2 \mu_2 \mu_3 (\mu_1-\mu_4)^2 \rho^4 \Big]}, \label{metric6dWeyl4}\\
G_{w\phi} &=&  \frac{c_2 \mu_2  (\mu_1 - \mu_4) \rho^2 (\mu_4^2 + \rho^2)(\mu_2 \mu_4 + \rho^2)(\mu_3 \mu_4 + \rho^2)}{ \mu_4 \Big[\mu_1 (\mu_2 \mu_4 + \rho^2)^2(\mu_3 \mu_4 + \rho^2)^2 - c_2^2 \mu_2 \mu_3 (\mu_1 -\mu_4)^2 \rho^4\Big]}, \label{metric6dWeyl5}\\
G_{ww} &=& \frac{\mu_2  \Big[ \mu_1  (\mu_2 \mu_4 + \rho^2)^2(\mu_3 \mu_4 + \rho^2)^2 + c_2^2 \mu_2 \mu_3 \mu_4^2 (\mu_1 - \mu_4)^2 \rho^2 \Big]}{\mu_4 \Big[\mu_1 (\mu_2 \mu_4 + \rho^2)^2(\mu_3 \mu_4 + \rho^2)^2 - c_2^2 \mu_2 \mu_3 (\mu_1 -\mu_4)^2 \rho^4\Big]}. \label{metric6dWeyl6}
\eea
The remaining components not related to these by symmetry vanish.
Note that setting $c_1 = c_2 = 0$ we get back the static seed metric~\eqref{seed}. The final conformal factor reads as:
\bea
\hspace{-0.7cm}
e^{2\nu} = k^2 \frac{(\mu_0 \mu_3 + \rho^2)(\mu_1 \mu_4 + \rho^2) \Big[M_{0,0} + c_1^2 M_{c_1,0} + c_2^2 M_{0,c_2} + c_1^2 c_2^2 M_{c_1, c_2}\Big]}{\mu_0(\mu_0 \mu_1+\rho^2)(\mu_0 \mu_2+\rho^2)(\mu_1 \mu_3+\rho^2)^2(\mu_2 \mu_4+\rho^2)(\mu_3 \mu_4+\rho^2)\prod_{i=0}^{4}(\mu_i^2+\rho^2)} \label{metric6dWeyl7}
\eea
where
\bea
M_{0,0} &=& \mu_1 \mu_3 (\mu_0 \mu_1 + \rho^2)^2 (\mu_0 \mu_2 + \rho^2)^2 (\mu_2 \mu_4 + \rho^2)^2 (\mu_3 \mu_4 + \rho^2)^2,  \label{metric6dWeyl8} \\
M_{c_1,0} &=&  \mu_0^2 \mu_1^2 \mu_2 (\mu_0-\mu_3)^2 \rho^2(\mu_2 \mu_4 + \rho^2)^2 (\mu_3 \mu_4 + \rho^2)^2,  \label{metric6dWeyl9}\\
M_{0,c_2} &=& - \mu_2 \mu_3^2 (\mu_1-\mu_4)^2 \rho^4 (\mu_0 \mu_1 + \rho^2)^2(\mu_0 \mu_2 + \rho^2)^2,  \label{metric6dWeyl10} \\
M_{c_1, c_2} &=& -\mu_0^2 \mu_1 \mu_2^2 \mu_3 (\mu_0-\mu_3)^2(\mu_1-\mu_4)^2 \rho^6 . \label{metric6dWeyl11}
\eea
The metric functions at this stage look somewhat complicated. However, as we will shortly see, these functions dramatically simplify when we express them in $(x,y)$ coordinates. To verify that the metric describes the singly spinning dipole black ring,  we first analyze the rod structure of the solution. This analysis is straightforward~\cite{Harmark:2004rm}. We find
\begin{figure}[t!]
\centering{
\includegraphics{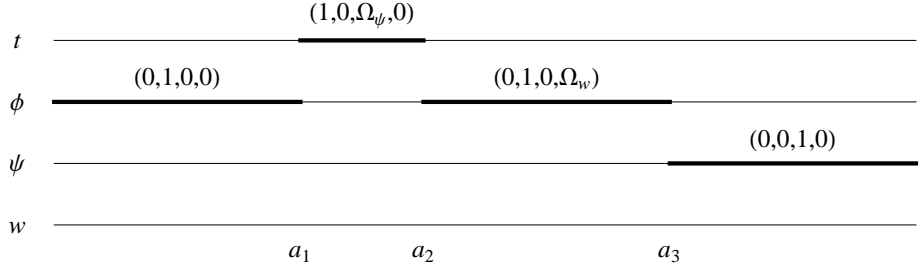}}
%\bigskip
\caption{The figure depicts sources for the six-dimensional lift of the dipole black ring. The direction for each rod is indicated. The points $a_0$ and $a_4$ are not indicated in the figure, as they no longer represent real turning points.}
\label{roddiagram1}
\end{figure}

\begin{itemize}
\item The semi-infinite rod $ z \in (-\infty, a_0]$ has direction $(0,1,0,0)$. The $G_{\psi \psi}$ component of the metric however diverges as $z \to a_0$ for general values of the BZ parameter $c_1$.
\item Similarly the rod $z \in [a_0, a_1]$ has direction $(0,1,0,0)$. Again the $G_{\psi \psi}$ component of the metric diverges as $z \to a_0$ for general values of the BZ parameter $c_1$.

Fortunately, the singularities in both $(-\infty, a_0]$ and $[a_0, a_1]$ rods  are completely removed by setting
\be
|c_1|= \sqrt{\frac{2(a_1 - a_0)(a_2 - a_0)}{(a_3 - a_0)}}.
\label{singremoval}
\ee
The condition~\eqref{singremoval} will be imposed from now onwards. Having imposed this condition we have effectively moved the negative density rod $z \in [a_0, a_1]$ from the $\psi$ direction to the $t$ direction and have canceled it against the segment $[a_0, a_1]$ of the rod $[a_0, a_2]$. In the process we have successfully generated rotation in the $\psi$ direction as is illustrated by the rod direction of the rod $[a_1,a_2]$.

\item The finite timelike rod $[a_1,a_2]$ has rod direction $(1,0,\Omega_\psi,0)$, where
\be
\Omega_\psi = \sqrt{\frac{(a_1 - a_0)}{2(a_2 - a_0)(a_3 - a_0)}}. \label{OmegaPsi}
\ee
\item The finite rod $[a_2,a_4]$ has direction $(0,1,0, \frac{2(a_4-a_2)}{c_2})$.
\item The finite rod $[a_4,a_3]$ has direction $(0,1,0, \frac{(a_4-a_1)c_2}{a_3-a_4})$.

Note that the directions of rods $[a_2,a_4]$ and $[a_4,a_3]$ are in general different. However, if we choose
\be
|c_2| = \sqrt{\frac{2(a_4 - a_2)(a_3 - a_4)}{(a_4 - a_1)}},
\label{rodalign}
\ee
then the two rod directions are identical. The condition~\eqref{rodalign}  will also be imposed from now onwards. Having imposed this condition we have effectively displaced the positive density rod $z \in [a_2, a_4]$ from the $w$ direction towards the $\phi$ direction and have mergerd it to the rod $[a_4, a_3]$. In the process we have successfully generated a dipole charge as illustrated by the rod direction $(0,1,0, \Omega_w)$ of the rod $[a_2,a_3]$, where
\be
\Omega_w = \sqrt{\frac{2(a_4-a_2)(a_4 - a_1)}{(a_3 - a_4)}}.
\label{Omegaw}
\ee
\item Finally, the semi-infinite rod $z \in [a_3, \infty)$ has rod direction $(0,0,1,0)$.
\end{itemize}
All of this discussion is succinctly illustrated in the rod diagram in figure~\ref{roddiagram1}.

\bigskip
To verify that the above described solution is indeed the singly spinning dipole black ring of~\cite{Emparan:2004wy} lifted to six dimensions, we now rewrite the metric in the ring coordinates $(x,y)$. First recall that the five dimensional fields for the dipole ring  take the form~\cite{Emparan:2004wy}
\bea
ds_5^2 &=& - \frac{F(y)}{F(x)}\left(\frac{H(x)}{H(y)}\right)^{1/3} \left( dt + C(\nu, \lambda) R \frac{1+y}{F(y)} d\psi\right)^2 \\
& &  + \frac{R^2}{(x-y)^2} F(x) \left(H(x)H(y)^2\right)^{1/3}\left[ - \frac{G(y)}{F(y) H(y)} d \psi^2 - \frac{dy^2}{G(y)}+ \frac{dx^2}{G(x)} + \frac{G(x)}{F(x)H(x)}d\phi^2 \right], \nn
\eea
where
\be
F(\xi) = 1 + \lambda \xi, \qquad G(\xi) = (1-\xi^2)(1 + \nu \xi), \qquad H(\xi) = 1- \mu \xi,
\ee
and
\be
 C(\nu, \lambda) = \sqrt{\lambda(\lambda - \nu)\frac{1+ \lambda}{1- \lambda}}.
\ee
The vector field $A$ and dilaton $\phi$ take the following expressions:
\be
A = C(\nu, - \mu) R \left(\frac{1+x}{H(x)} \right) d \phi, \qquad e^{- \phi} = \left(\frac{H(x)}{H(y)}\right)^{\sqrt{2/3}}.
\ee
The six-dimensional lift can be readily obtained using the KK ansatz~\eqref{metric6d}. We get
\bea
ds_6^2 &=& - \frac{F(y)}{F(x)}\left( dt + C(\nu, \lambda) R \frac{1+y}{F(y)} d\psi\right)^2 + \frac{H(x)}{H(y)}\left(d w + C(\nu, - \mu)R \frac{1+x}{H(x)} d\phi\right)^2  \label{metric6dxy}
 \\
& &  + \frac{R^2}{(x-y)^2} F(x) H(y)\left[- \frac{G(y)}{F(y) H(y)} d \psi^2 - \frac{dy^2}{G(y)}+ \frac{dx^2}{G(x)} + \frac{G(x)}{F(x)H(x)}d\phi^2 \right]. \nn
\eea
Note in particular that in six dimensions all fractional powers of metric functions have disappeared.

Our aim now is to show that the metric in Weyl coordinates~\eqref{metric6dWeyl1}--\eqref{metric6dWeyl11} is identical to the metric~\eqref{metric6dxy}. To establish this we follow the procedure of~\cite{Harmark:2004rm} and appendix A.2 of~\cite{Elvang:2007rd}. The coordinate transformation from Weyl $(\rho, z)$ to ring coordinates $(x,y)$ is
\be
\rho = \frac{R^2 \sqrt{-G(x)G(y)}}{(x-y)^2}, \qquad z = \frac{R^2(1-xy)[2 + \nu(x+y)]}{2(x-y)^2}.
\ee
The rod end points are related to parameters $\nu, \lambda, \mu$ as
\be\label{constants}
a_0 = \frac{R^2}{2} \alpha, \quad a_1 = - \frac{R^2}{2}\nu, \quad a_2 = \frac{R^2}{2}\nu, \quad a_3 = \frac{R^2}{2}, \quad a_4 = \frac{R^2}{2}\beta.
\ee
Here $-\infty < \alpha \le - \nu$ and $\nu \le \beta < 1$ are constants to be determined. After a straightforward calculation we see that with the choice
\be
\alpha = \frac{\nu(1+\lambda)- 2 \lambda}{(1-\lambda)}, \qquad  \beta = \frac{\nu(1-\mu)+ 2 \mu}{(1+\mu)}, \qquad k^2 = \frac{(1 + \mu) (1-\lambda)}{(1-\nu)^2}, \label{params}
\ee
all metric components match. The conditions $-\infty< \alpha \le - \nu$ and $\nu \le \beta < 1$ and the ordering of the rod end points imply that
\be
0 < \nu \le \lambda < 1 \qquad \mbox{and} \qquad 0\le \mu < 1.
\ee
We therefore have also recovered the correct bounds on the parameters of the singly spinning dipole black ring. The periodicities of angular coordinates are $\Delta \phi = \Delta \psi = 2 \pi k$. With $k$ given in \eqref{params}, these periodicities also agree with the results of~\cite{Emparan:2004wy}. Thus we have shown that the solution~\eqref{metric6dWeyl1}--\eqref{metric6dWeyl11} is identical to~\eqref{metric6dxy}.

%%%%%%%%%%%%%%%%%%%%%%%%%%%%%%%%%%%%%%%
\section{Discussion and Outlook}
Using GL(4,~R) integrability structure of the  Einstein-Maxwell-dilaton theory with dilaton coupling $a = 2 \sqrt{2/3}$, we presented an inverse scattering construction of a singly spinning dipole black ring solution. The finite rod in the Kaluza-Klein direction $w$ plays the key role in our construction. Recall that finite rods in Kaluza-Klein directions correspond to KK bubbles~\cite{Emparan:2001wk}. Therefore, loosely said, KK bubble is converted into magnetic dipole in our inverse scattering construction.

The novel construction presented above can in principle be generalized to construct the doubly spinning dipole black ring and/or the most general black ring in this theory. We emphasize once again that this situation is in contrast with the previous approaches, e.g.,~\cite{Yazadjiev:2006ew}, which rely on a specific SL(2,~R) $\times$ SL(2,~R) structure, and cannot be extended in a natural way to construct, say, the doubly spinning dipole black ring. Our approach does not rely on a SL(2,~R) $\times$ SL(2,~R) structure, but rather uses the full GL(4,~R) symmetry of the theory.

The approach presented above has some resemblance to the approach taken in~\cite{DGGH}, where the first static self-gravitating loop of string was described. There, starting with the product of five dimensional Euclidean Schwarzschild and time, and performing a twisted KK reduction, a static ring solution was obtained. The solution is regular, but the horizon of the ring in that case is degenerate (extremal) with zero area. The solution is balanced not by rotation but by a background field. Due to the presence of the background field, the solution is not asymptotically flat, instead it approaches a fluxbrane.
In our approach, since we mix the $w$ and $\phi$ directions via the inverse scattering method, we generate the
five-dimensional magnetic charge without the need to do a twisted KK reduction. Thus, we obtain an asymptotically flat solution in five-dimensions instead of a fluxbrane. A related technical point that is worth mentioning is that in our generation of the singly spinning dipole ring it is not necessary to perform any mixing of the Killing coordinates to obtain the desired asymptotic structure.

Finally, in the process of constructing the dipole black ring we also found its six dimensional rod diagram description. A common picture seems to be that black hole solutions are uniquely characterized by their charges and rod diagrams~\cite{Tomizawa:2009tb}. We expect that the rod diagram description of the dipole black ring presented above will play an essential role in the uniqueness proofs within the described dilaton theory. Some progress in this direction was already made in~\cite{Yazadjiev:2011fg}.

The new procedure implemented in this paper sets the building blocks for the construction of more general and exotic black hole solutions with dipole charges. One could add dipole charges to e.g. the bicycling black ring of~\cite{Elvang:2007hs}, the di-ring of~\cite{Iguchi:2007is} or a further rotation to the dipole black ring. In this work we have not attempted to construct the doubly spinning dipole solution, or the most general black ring solution, but we hope to return to these problems elsewhere.

%%%%%%%%%%%%%%%%%%%%%%%%%%%%%%%%%%%%%%%
\subsection*{Acknowledgements}
\begin{sloppypar}
We thank Joan Camps, Pau Figueras and Oscar Varela for discussions. We specially thank Roberto Emparan for comments on a draft version and for bringing reference \cite{DGGH} to our attention as well as the organizers of the `Gravity' workshop, 17-29 July 2011, held at Centro de Ciencias Pedro Pascual, Benasque, Spain where this work was completed. JVR is supported by {\it Funda\c{c}\~ao para a Ci\^encia e Tecnologia} (FCT)-Portugal through contract no. SFRH/BPD/47332/2008. AV was supported by IISN Belgium (conventions 4.4511.06 and 4.4514.08) and by the Belgian Federal Science Policy Office through the Interuniversity Attraction Pole P6/11.
\end{sloppypar}

%%%%%%%%%%%%%%%%%%%%%%%%%%NEW BIB%%%%%%%%%%%%%%%%%%

\end{document}